\documentclass[aps,prb,twocolumn]{revtex4}

\usepackage{amsmath}
\usepackage{amssymb}
\usepackage{amsfonts}
\usepackage{amsthm}
\usepackage{graphicx}
\usepackage{hyperref}
\usepackage{color}
\usepackage{subfigure}
\usepackage{soul}

\newcommand{\tr}{\operatorname{Tr}}
\newcommand{\ket}[1]{\left | #1 \right \rangle}
\newcommand{\bra}[1]{\left \langle #1 \right |}

\newcommand{\etal}{\emph{et al.}}

\begin{document}

\title{Experimental realisation of Shor's quantum factoring algorithm using qubit recycling}
\author{Enrique Mart\'{i}n-L\'{o}pez}
\altaffiliation{These authors contributed equally to this work}
\author{Anthony Laing}
\altaffiliation{These authors contributed equally to this work}
\author{Thomas Lawson}
\altaffiliation{These authors contributed equally to this work}
\author{Roberto Alvarez}
\author{Xiao-Qi Zhou}
\author{Jeremy L. O'Brien}
\email{Jeremy.OBrien@bristol.ac.uk}
\affiliation{Centre for Quantum Photonics, H. H. Wills Physics Laboratory \& Department of Electrical and Electronic Engineering, University of Bristol, Merchant Venturers Building, Woodland Road, Bristol, BS8 1UB, UK}

\begin{abstract}
Quantum computational algorithms exploit quantum mechanics to solve problems exponentially faster than the best classical algorithms \cite{fe-ijtp-82-467,de-prsla-400-97,nielsen}.  Shor's quantum algorithm \cite{sh-conf-94-124} for fast number factoring is a key example and the prime motivator in the international effort to realise a quantum computer \cite{la-nat-464-45}.  However, due to the substantial resource requirement, to date, there have been only four small-scale demonstrations \cite{va-nat-414-883,lu-prl-99-250504,la-prl-99-250505,po-sci-325-1221}.  Here we address this resource demand and demonstrate a scalable version of Shor's algorithm in which the $n$ qubit control register is replaced by a single qubit that is recycled $n$ times: the total number of qubits is one third of that required in the standard protocol \cite{pa-prl-85-3049, mo-LNCS-1509-174, ParkerPlenioNote}.  Encoding the work register in higher-dimensional states, we implement a two-photon compiled algorithm to factor $N=21$.  The algorithmic output is distinguishable from noise, in contrast to previous demonstrations.  These results point to larger-scale implementations of Shor's algorithm by harnessing scalable resource reductions applicable to all physical architectures.
\end{abstract}

\maketitle

Shor's factoring algorithm consists of a quantum order finding algorithm, preceded and succeeded by various classical routines.  While the classical tasks are known to be efficient on a classical computer, order finding is understood to be intractable classically. However, it is known that this part of the algorithm can be performed efficiently on a quantum computer.  To determine the prime factors of an odd integer $N$, one chooses a co-prime of $N$, $x$. The order $r$ relates $x$ to $N$ according to $x^r\!\!\!\! \mod N =1$, and can be used to obtain the factors, given by the greatest common divisor $\gcd(x^{\frac{r}{2}}\pm1, N)$.

\begin{figure*}[t]
\centering
\includegraphics[width=\textwidth]{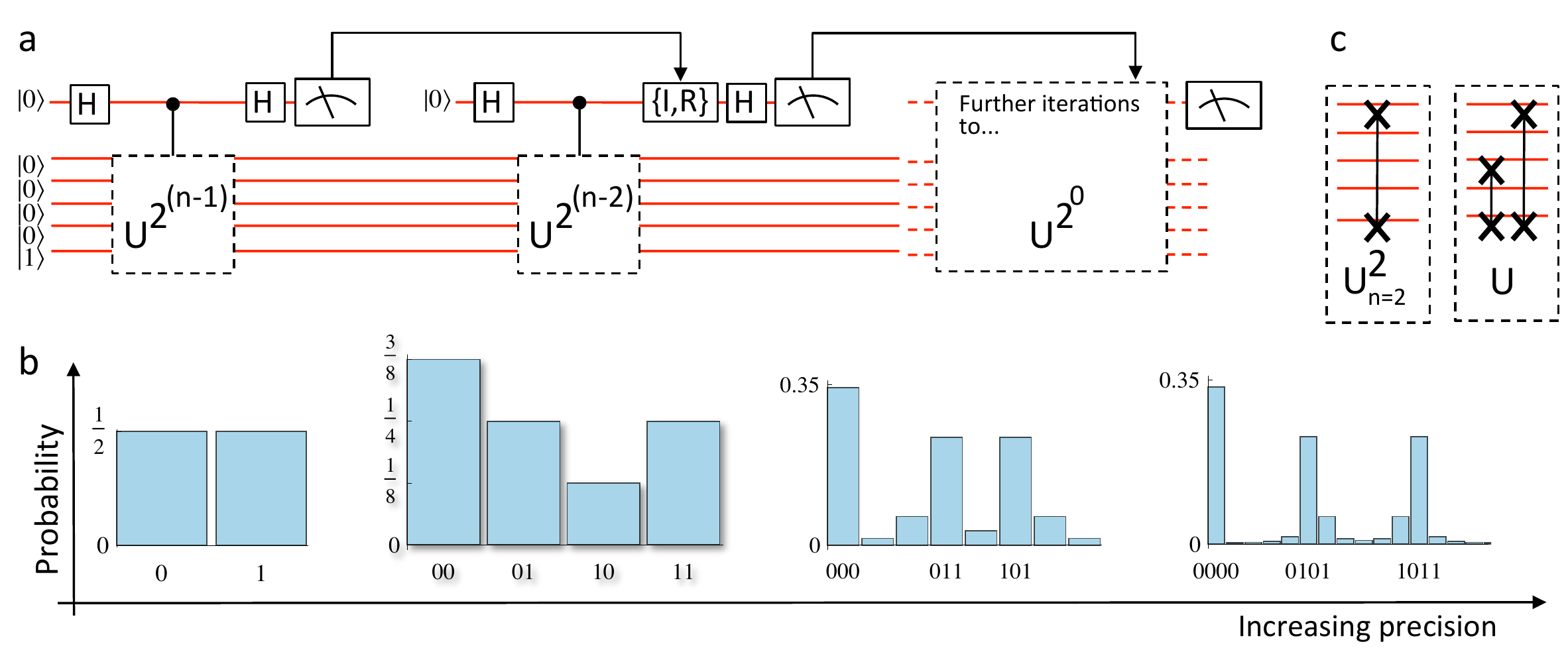}
\caption {The iterative order finding algorithm for factoring $21$.
{\bf a}, Measurement of the control qubit after each controlled unitary gives the next most significant bit in the output and the outcome is fed forward to the iterated (semi-classical) Fourier transform, which applies either the identity operation $I$ or the appropriate phase gate $R$, prior to the Hadamard $H$. {\bf b}, As the number of iterations increases the precision increases. {\bf c}, For two bits of precision the controlled unitary operations can be constructed with this arrangement of controlled-swap gates.}
\label{fgTheory}
\end{figure*}

The quantum order finding circuit involves two registers: a work register and a control register.  In the standard protocol, the work register performs modular arithmetic with $m=\lceil \log_{2}N \rceil$ qubits, enough to encode the number $N$, and the $n$ qubit control register provides the algorithmic output, with $n$ bits of precision.  

Measuring the control register in the computational basis will yield a result of $k 2^{n}/r$ where $k$ is an integer between $0$ and $r-1$, with the value of $k$ occurring probabilistically.  Dividing the result by $2^{n}$ gives the first $n$ bits of ${k}/{r}$ and $r$ may be found with classical processing, using the continued fraction algorithm.  For large $n$, and a perfectly functioning circuit, the output probability distribution of the control register is a series of well defined peaks at values of $k 2^{n}/r$ (Fig.~\ref{fgTheory}b). (See \emph{Appendix} for details.)

Here we implement an iterative version of the order finding algorithm \cite{pa-prl-85-3049, mo-LNCS-1509-174}, in which the control register contains only a single qubit which is recycled $n$ times, using a sequence of measurement and feed-forward operations, with each step providing an additional bit of precision (Fig.~\ref{fgTheory}a).  Reducing the number of qubits in quantum simulations and quantum chemistry has been achieved with recursive phase estimation \cite{as-sci-309-1704, la-nchem-2-106, ve-jcp-133-194106, wh-mp-109-735}, while ground state projections have been demonstrated by exploiting similar techniques in NMR \cite{li-sr-1-735}.

The iterative version of the order finding algorithm, displayed in Fig.~\ref{fgTheory}a, is closely related to the semi-classical picture of the quantum Fourier transform \cite{gr-prl-76-3228}.  Rather than performing a Fourier transformation across all control register qubits simultaneously, it is sufficient to measure the coherence between computational basis states of individual qubits in the control register (from least significant bit to greatest significant bit) deciding on the phase coherence of the next measurement depending on the previous result.  In fact, the control register need not contain more than one qubit at any one time.  This register begins in a product state and all unitaries controlled from the (i)th control qubit are performed before those controlled from the (i+1)th control qubit.  This means that all operations on the (i)th qubit can be performed before the (i+1)th qubit is initialised and a single control qubit can be re-used, or recycled, with the state of the work register iteratively updated at each stage.

To guarantee a good level of precision in $k/r$, a full scale implementation of Shor's algorithm requires $n$ to be $log(N^2)\lesssim n\lesssim log(2N^2)$.  So for full scale implementations, qubit recycling reduces the total number of qubits required from $\lceil 3 \log_{2}N \rceil$ to $\lceil \log_{2}N \rceil+1$; the only penalty is a polynomial increase in computation time, while the exponential speedup is retained---\textit{i.e.}~it is scalable.  In general, saving in qubits can be more than $2/3$ if more control qubits are required, or less than $2/3$ in smaller proof of principle demonstrations such as this.

As the number of control qubits, or iterations, $n$ is reduced, in general, the precision is reduced and the $k$ peaks in the probability distribution become smeared, as shown in Fig.~\ref{fgTheory}b.  However, in the special case where the order is a power of two and $r=2^{p}$, the peaks correspond exactly to the logical states of $p$ qubits 
\cite{FirstNote} 
such that the output is equivalent to that of an incoherent mixture of $p$ qubits (any additional control qubits remain unentangled throughout the algorithm and simply encode the trailing zeros in the uniform distribution).  Factoring $N=15$ gives either order 2 or order 4 for each of its co-primes \cite{va-nat-414-883,lu-prl-99-250504,la-prl-99-250505,po-sci-325-1221}; independent verification of entanglement is therefore required \cite{lu-prl-99-250504,la-prl-99-250505}.
For this reason we focus on $N=21$ with the co-prime $x=4$ to give order
\cite{SecondNote} 
$r=3$.

To two bits of precision the expected outcomes ${00}$, ${01}$, ${10}$ and ${11}$ occur with probabilities $\frac{3}{8}$, $\frac{1}{4}$, $\frac{1}{8}$ and $\frac{1}{4}$, respectively (Fig \ref{fgTheory}b).  Quantum interference in the Fourier transform is constructive for states contributing to the $00$ term and boosts its probability of observation to three times that of the probability for observing the $10$ term, which experiences destructive quantum interference among its contributory states.  Decoherence in the Fourier transform would degrade the contrast between these terms.  For states contributing to the $01$ and $11$ terms, the Fourier transform imparts phases to equalise the probability of observing these outcomes.  Therefore, the underlying periodic structure apparent in the two qubit probability distribution is susceptible to decoherence, and therefore distinguishable from noise (See \emph{Appendix} for details).

\begin{figure*}[t]
\centering
\includegraphics[width=\textwidth]{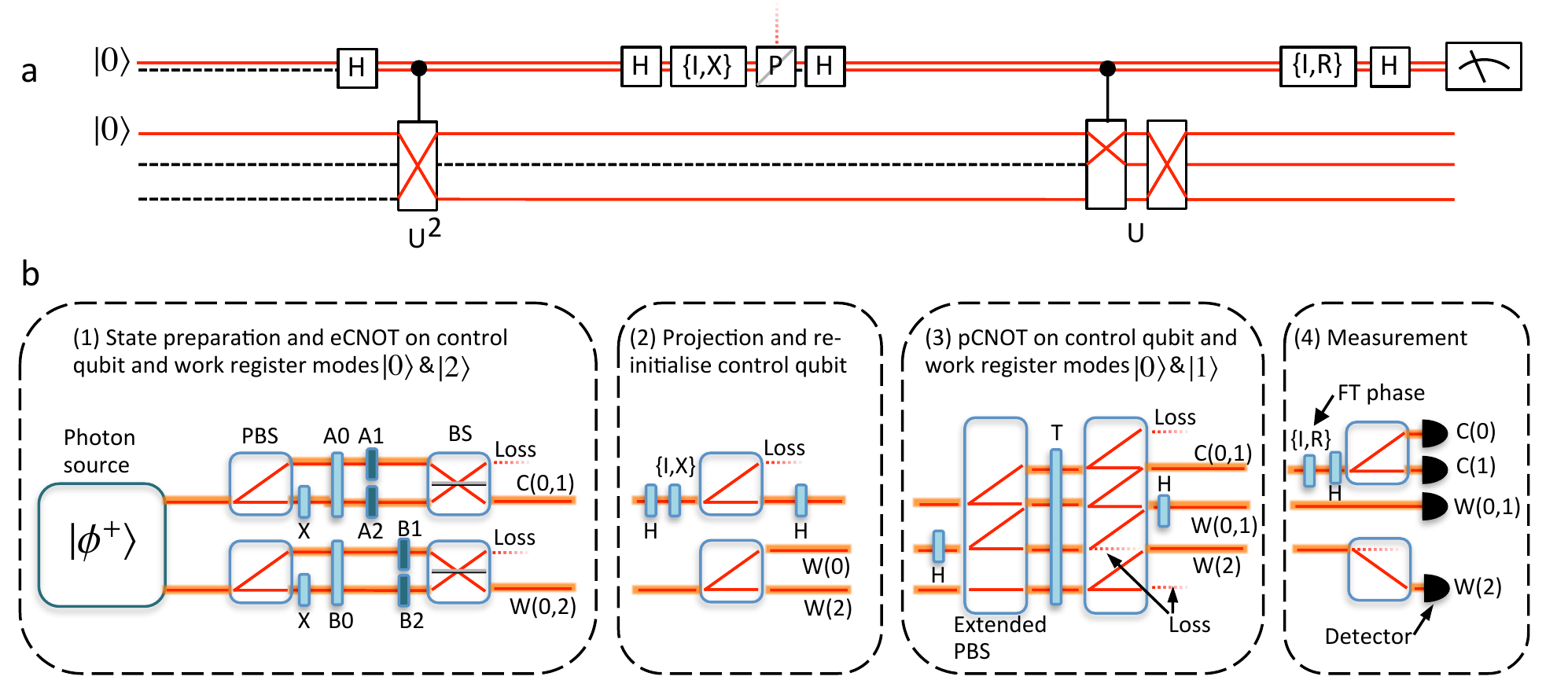}
\caption{Compiled iterative order finding algorithm. {\bf a}, Idealised compiled circuit diagram.  The first controlled unitary is implemented as a CNOT acting on work modes $0$ and $2$; the final controlled unitary uses a CNOT on work modes $0$ and $1$ followed by an uncontrolled swap on work modes $0$ and $2$.  $P$ indicates a projection onto the computational $0$, preceded by either a bit flip gate $X$, or the identity operation $I$; $H$ denotes the Hadamard gate. The iterative Fourier transform includes the rotation $R= \ket{H}\bra{H} - i \ket{V}\bra{V}$ when the first projection is made on to the computational $1$ state (See \emph{Appendix} for details).  {\bf b}, Schematic of the experimental circuit (See \emph{Methods} for details).}
\label{fgCirc}
\end{figure*}

We implement a scalable iterative quantum algorithm with a compiled version of the quantum order finding routine where the circuit is constructed for a particular factoring case (here $N=21$ and $x=4$) admitting an experimentally tractable implementation, as shown in Fig.~\ref{fgCirc}a.  There are several steps to compilations, common to previous demonstrations \cite{va-nat-414-883,lu-prl-99-250504,la-prl-99-250505,po-sci-325-1221,compiled}.  Firstly, for $N=21$ and $x=4$, unitaries and their decompositions can be calculated explicitly, as can the full state evolution; secondly, redundant elements of these unitaries are omitted.  Finally, and specific to our own demonstration, since only $3$ of the possible $2^{5}$ levels of the conventional $5$ qubit work register are ever accessed, a single qutrit is used for the work register, instead of $5$ qubits, to realise a hybrid qubit-qutrit system
\cite{ThirdNote} .
(See \emph{Appendix} for details).

The controlled unitaries that apply the function $f(x,a,N)=x^{a}\pmod{N}$ (where $a$ is a control register computational basis state) to the work register may be realised with a sequence of controlled swaps.  For our two qubit control register, the single swap of $U^{2}$ is implemented with a controlled-NOT (CNOT) gate; $U^{1}$ is realised with two swaps, the first of which is a CNOT gate, while it is sufficient for the second swap to be uncontrolled.
(See \emph{Appendix} for details).

Our scheme therefore requires two consecutive photonic CNOT gates---something that has not previously been demonstrated---acting on qubit subspaces of the qutrit.  In our experimental implementation we use the iterative approach with post-selection in place of measurement and feed-forward; measurement and feed-forward operations have been achieved in the context of cluster state quantum computing with photons \cite{pr-nat-445-65}.  The quantum order finding circuit of Fig.~\ref{fgCirc}a was experimentally constructed using the optical circuit shown schematically in Fig.~\ref{fgCirc}b. Realising two consecutive CNOT gates on two photons was achieved by using an entanglement driven CNOT gate (eCNOT) \cite{Zhou+Ralph2010}, followed by a post-selected CNOT gate (pCNOT) \cite{ra-pra-65-062324,ho-pra-66-024308,ob-nat-426-264,ob-prl-93-080502} (pCNOT gates cannot be used in series without the addition of ancilla photons).  The circuit was constructed with Jamin-Lebedeff polarisation interferometers in a calcite beam displacer architecture, chosen to provide interferometric stability.  Photons were generated with a polarisation entangled spontaneous parametric down conversion source\cite{kw-pra-60-773} (See \emph{Methods} for further experimental details).

\begin{figure}[t]
\centering
\includegraphics[trim=0 0 0 0, clip, width=\columnwidth]{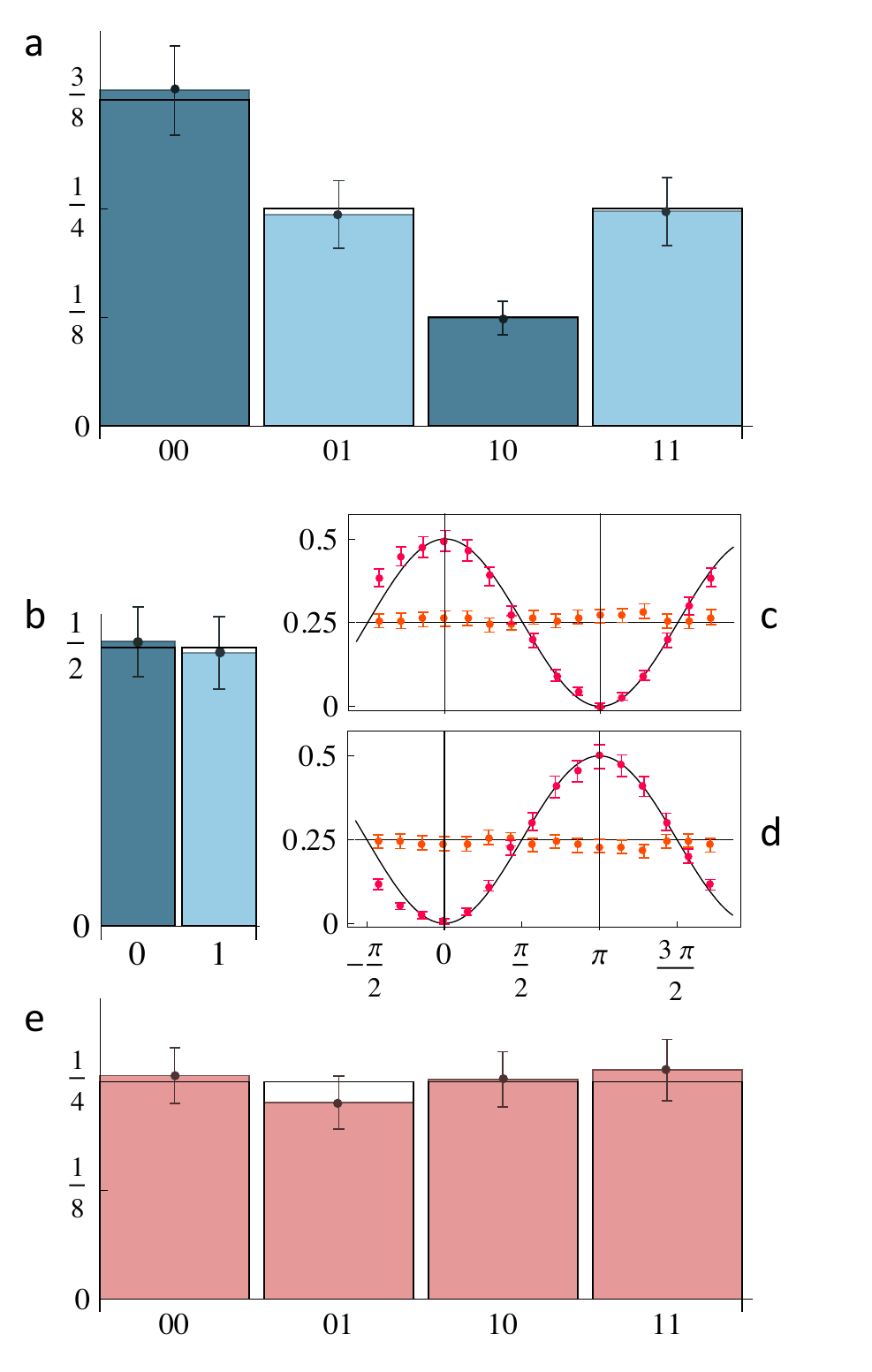}
\caption{Demonstration of order finding. Ideal probability distributions are plotted in solid black lines.
{\bf a}, The two-bit output probability distribution for the order finding algorithm.
{\bf b}, Output from the first iteration of the order finding algorithm.
{\bf c} \& {\bf d}, Probability distributions for 16 phases of the Fourier transform, as described in the text, for control qubit detectors 0 and 1 respectively.
{\bf e}, The distribution obtained from experimental simulation of decoherence in the Fourier transform.
Data were corrected for the measured difference in total coupling efficiency between the work register detectors.}
\label{fgRes}
\end{figure}

The correct algorithmic output from the quantum order finding circuit for factoring $N=21$ is confirmed by the data shown in Fig.~\ref{fgRes}a.  The two-qubit control register output probabilities for ${00}$, ${01}$, ${10}$ and ${11}$ were measured and found to have a fidelity of  $99 \pm 4 \%$
\cite{FourthNote} 
with the ideal probabilities $\frac{3}{8}$, $\frac{1}{4}$, $\frac{1}{8}$ and $\frac{1}{4}$, respectively.  The distribution on the first qubit, which determines the second bit in the total probability distribution, was measured by comparing total control register counts, heralded by the $W(2)$ detector, for each setting of the first qubit projector (stage (2) of the circuit in Fig.~\ref{fgCirc}b) and found to be uniform with fidelity $99 \pm 5 \%$ (Fig.~\ref{fgRes}b).

Since the non-uniformity of the distribution of Fig.~\ref{fgRes}a arises from the second iteration of the algorithm when the first iteration gives output zero, we now focus on this case. Analysis of the circuit and algorithm reveal a critical dependence on decoherence, particularly in the pCNOT and the Fourier transform phase:  phase instability drives the output toward a uniform probability distribution.  We confirmed this analysis experimentally, as described below. 

With stage $2$ of the circuit set so that the first qubit is projected onto the computational zero state, Figs.~\ref{fgRes}c and \ref{fgRes}d show $16$ probability distributions for the correct phase setting (\emph{i.e.}~0) to implement the semiclassical Fourier transform, and 15 incorrect settings of this phase
\cite{FourthNoteB}.
When heralded by $W(0,1)$, the detector that does not distinguish between the $0$ and $1$ states of the work qutrit, the control qubit should be in a maximal mixture and therefore insensitive to this phase---\emph{i.e.}~give a flat line response. In contrast, since mode $2$ of the qutrit is not involved in the final CNOT gate, the control qubit is not entangled with it.  Therefore, when heralded by the work qutrit being in the $2$ state, the control photon should be in a pure state and should exhibit maximum sensitivity to the phase in the Fourier transform---\emph{i.e.}~a unit visibility fringe.  Taking into account the relative probability amplitudes in the ideal output state, the ideal plot of probability distributions would show two sinusoidal curves of full visibility (from detectors $C(0) \& W(2)$ and $C(1) \& W(2)$) each bisected by a flat line (from detectors $C(0) \& W(0,1)$ and $C(1) \& W(0,1)$).  For the $8$ probability distributions between two settings of the Fourier transform phase, $0$ and $\pi$ (indicated by black vertical lines) we find an average fidelity between this situation and our data of $91.6 \pm 0.6 \%$.  

The data in Figs.~\ref{fgRes}c and \ref{fgRes}d can be used to show the sensitivity of the circuit to decoherence: Integrating the probability distribution over the range 0 to $\pi$ of the Fourier transform phase simulates the effect of phase instability.   The red distribution plotted in Fig. \ref{fgRes}e shows the near uniform distribution that results from this procedure, confirming susceptibility of this circuit to decoherence. 

\section*{Methods}
\noindent\textbf{Photon source:} Entangled photon pairs, spectrally degenerate at 808 nm, were generated in a Type 1 spontaneous parametric down conversion source, with a 404 nm CW laser was focused to a 40 $\mu$m waist in a pair of crossed $\text{BiB}_3\text{O}_6$  (Bismuth Triborate) non linear crystals \cite{kw-pra-60-773}.  Photons were spectrally filtered with high transmission interference filters of FWHM 3nm, then collected into polarisation maintaining optical fibres (PMFs).  PMFs would normally decohere the polarisation of photons that are not aligned with the slow or fast axis of the fibre, as is the case for photons that are entangled.  Our fibres were cut at the midpoint and spliced together with a 90 degree twist such that the slow axis in the first length was aligned with the fast axis in the 2nd length.  While this modified fibre imparts an unknown phase shift between the two polarisations, their coherence is preserved.  The unknown unitary is pre-compensated with wave plates in the source. State tomography of the photon source, which drives the eCNOT gate, revealed a highly entangled state with fidelity $96.9\pm 0.2\%$ to the corresponding Bell state 
\cite{FifthNote}.

\noindent\textbf{Optical Circuit:} The optical circuit of Fig. \ref{fgCirc}b was experimentally constructed using an architecture of calcite beam displacers (BD), which separate the ordinary and extraordinary polarisations and can be used to form very stable Jamin-Lebedeff interferometers---polarisation interferometers with parallel light-paths that provide interferometric stability.  The PBSs and Extended PBSs in Fig \ref{fgCirc} were directly implemented with a single BD.  The eCNOT gate \cite{Zhou+Ralph2010} requires two non polarising $50\%$ reflectivity beam splitters (BS in Fig \ref{fgCirc}) the unitary operation of which was constructed with four BDs and wave plates.  The action of polarisers A1 and A2 was realised using beam stops after the first BD in the BSs.  The operation of the circuit is as follows:

\emph{Stage 1 in Fig.~\ref{fgCirc}b}:
The experimental control and work registers are initialised within the eCNOT gate by respectively configuring the $A0$ and $B0$ wave plates to output the desired states as if each of their inputs were the computational $\ket{0}$: $A0$ is set to implement the Hadamard and $B0$ implements the Identity operation.  The eCNOT gate is driven by pre-entanglement \cite{Zhou+Ralph2010} from the polarisation entangled SPDC source in the state $\ket{1_{H,U}}\ket{1_{H,L}} + \ket{1_{V,U}}\ket{1_{V,L}}$ (where $H/V$ denotes horizontally/vertically polarised light and U/L denotes upper/lower path) which is then converted to path entanglement with polarisation beam splitters (PBS) and polarisation flips (X).  After combing the two double-rails on non-polarising beam splitters (BS) and post selecting on the cases where photons emerge in the two lower paths, the $2\times2$ transition matrices of the optical elements $\{A1, A2, B1, B2\}$ combine as $A1\otimes B1+ A2 \otimes B2$: choosing a vertical polariser for $A1$, a horizontal polariser for $A2$, the Identity operation for $B1$, and a polarisation flip for $B2$, implements the CNOT gate logic on the initialised states.  In its general form, the eCNOT gate can perform any controlled unitary operation (by choosing appropriate optical elements for $A1, A2, B1,$ and $B2$), and the addition of a KLM-like teleportation scheme \cite{kn-nat-409-46} allows the gate to work with non separable states.

The polarisation modes within the control spatial mode correspond to the qubit computational states indicated by $C(0,1)$; at this point the polarisation modes within the work spatial mode correspond to the $\ket{0}$ and $\ket{2}$ qutrit states, indicated by $W(0,2)$.

\emph{Stage 2 in Fig.~\ref{fgCirc}b}: The control qubit is projected onto one of the computational states, dependent upon whether $I$ or $X$ is performed before the upper PBS.  The lower PBS introduces the third mode for the work register so that the $\ket{0}$ and $\ket{1}$ states are polarisation encoded in the upper spatial mode of the work register (though at this stage the $\ket{1}$ state has zero probability amplitude, i.e. vacuum) while the lower spatial mode contains only one polarisation and corresponds to the $\ket{2}$ state.

\emph{Stage 3 in Fig.~\ref{fgCirc}b}: The pCNOT gate relies on photonic quantum interference tuned by the half wave plate $T$ which is set to $62.5^{\circ}$.  Successful operation is heralded when one photon is present in the control modes and one photon is present in the work modes.  Here, further balancing loss is introduced into the $W(2)$ mode.  The output from $W(2)$ and the usual pCNOT work loss mode share the same spatial mode but different polarisations.  The entangling capability of the pCNOT gate was tested with a Bell inequality violation (while in situ) recording a CHSH value of  $2.67 \pm 0.01$  (violating the classical limit of 2 by $55$ standard deviations).

\emph{Stage 4 in Fig.~\ref{fgCirc}b}:  The control qubit is assigned a phase according to the projector in the first iteration, allowing implementation of the semi-classical Fourier transform.  The control qubit states are individually projected and provide the order finding results.  At the final stage, the work qutrit plays no role in providing order finding information (other than to herald the control qubit) so individual computational states may be traced out in detection.  The polarisations of the upper work spatial mode are not distinguished, but the remaining work mode is; these two cases provide a useful method to confirm correct circuit operation.

\medskip
\begin{acknowledgments}
\vspace{-4 mm}
\noindent We thank
Stephen Bartlett, Richard Jozsa, Gary McConnell, Tim Ralph, Terry Rudolph and Pete Shadbolt for helpful discussions. This work was supported by the Engineering and Physical Sciences Research Council (EPSRC), the European Research Council (ERC), PHORBITECH and the Centre for Nanoscience and Quantum Information
(NSQI). J.L.O'B. acknowledges a Royal Society Wolfson Merit Award. 
\end{acknowledgments}

\medskip
\section*{Author contributions}
\vspace{-4 mm}
\noindent
Theory was developed by TL and AL.
Theory was mapped to experimental circuit by AL, TL, EML, XZ, and JOB.
Experiment was performed by EML, TL, AL, RA, and XZ.
Data was analysed by AL, EML, TL, XZ, and JOB.
Manuscript was written by AL, TL, EML, XZ, and JOB.
Project was supervised by AL and JLO'B.

\appendix
\begin{widetext}
\vspace{10 pt}
\section{Supplementary Information}

\noindent\textbf{Standard protocol operation:}
The quantum order finding circuit involves two registers: a work register and a control register.  In the standard protocol, the work register performs modular arithmetic with $m=\lceil \log_{2}N \rceil$ qubits, enough to encode the number $N$, and the $n$ qubit control register provides the algorithmic output, with $n$ bits of precision.  The control register, $\ket{c}$, starts in a superposition over all logical states, $\ket{c_{ini}}=\sum^{2^{n}-1}_{a=0}\ket{a}$.  The states of the work register, $\ket{w}$, are then computed as a function of those in the control register $\{\ket{a}\}$, the coprime $x$, and the number to be factored $N$, $\ket{w_{a}}=\ket{f(x,a,N)}=\ket{x^{a}\pmod{N}}$, to produce a highly entangled state, $\ket{c,w}=\sum^{2^{n}-1}_{a=0}\ket{a}\ket{x^{a}\pmod{N}}$.  The condition $x^{r}=1$ induces a periodicity in the work register so that the logical states of the control register can be grouped into $r$ sets as $\sum^{r-1}_{j=0} (\sum^{\lceil (2^{n}-j)/r \rceil}_{i=0} \ket{i r+j}) \ket{x^{j}}$ (states are not normalised here and throughout).  Crucially, the periodicity in control register states is exploited through a Fourier transformation, which transforms a function with period $r$ into a new function with period $2^{n}/r$.\\

\noindent\textbf{Compiling the algorithm:}
The circuit schematic shown in Fig.~2a is designed to perform factoring for the specific case of $N=21$ and $x=4$.  Controlled unitary operations perform $f(x,a,N)$ on the work register for each value of $a$ in the control register; for $N=21$ and $x=4$, the unitaries, their decompositions, and the full state evolution, can be calculated explicitly by hand and redundant elements of these unitaries are omitted (see next section for further details).  In the standard qubit encoding of Fig.~1a, the work register requires $5$ qubits to represent $N=21$ in binary.  However, only three (of the possible $2^{5}$) work register states are ever accessed since the function $f(x,a,N)=x^{a}\pmod{N}$ is a periodic repetition of $\{4^{0}=1,4^{1}=4,4^{2}=16\}$.  We implement the work register using a single qutrit, taking advantage of the further degrees of freedom available in photon path and polarisation encoding.  The qutrit levels represent  the active sates $\{1,4,16\}$ with the qutrit labels taking the $\log_{4}$, of these values.  This replaces the $5$ qubits shown in Fig.~1a.\\

\noindent\textbf{Unitary decomposition and algorithm operation:}
The function $f(x,a,N)=x^{a}\pmod{N}$ is carried out on the work register through a sequence of controlled unitary transformations $\{U^{2^{j-1}}\}$, with each unitary controlled from the (j)th control qubit (where $j=1$ is the final qubit, giving the most significant bit).   With the the work register initialised in the $\ket{w_{ini}}=\ket{1}$ state, $U^{1}$ should perform the mapping $x^{i}\rightarrow x^{i+1}\pmod{N}$.  Therefore, $f(x,a,N)$ is realised through a sequence of controlled permutations, and each controlled permutation may be realised with a sequence of controlled swaps.  For $N=21$ and $x=4$ with two control bits, $U^{2} : \{ 1\rightarrow16, 4\rightarrow1, 16\rightarrow4 \}$ is implemented first, followed by $U^{1} : \{ 1\rightarrow4, 4\rightarrow16, 16\rightarrow1 \}$.  It is shown in Fig.~2a that, since the standard work register starts in the state $\ket{1}$, only the mapping $\ket{1}\rightarrow\ket{16}$ from $U^{2}$ is required; with our relabelling for the work qutrit, the work register initialised as $\ket{0}$ should be mapped to $\ket{2}$, which is realised with a CNOT gate acting on a qubit subspace of the qutrit: $\ket{0}\leftrightarrow\ket{2}$.  $U^{1}$ may be performed with two swaps: $0\leftrightarrow1$ followed by $0\leftrightarrow2$.  The first of these swaps is implemented as a CNOT, while an uncontrolled swap, equivalent to a relabelling of two modes of the qutrit, is sufficient for the second (see below for further details).

The (still unnormalised) state of the qubit-qutrit system shown in Fig.~2a before the first projection is $\ket{c_{b}, w_{t}}_{1}=\ket{0}(\ket{0}+\ket{2})+\ket{1}(\ket{0}-\ket{2})$, so that selecting either the identity operation $I$ or the bit flip $X$ before the polariser $P$ post selects on to the first or second term respectively.  Ideally, there is an equal probability of observing $\ket{0}$ or $\ket{1}$ in the control register, which corresponds to the least significant digit in the final output.  With $\ket{0}$ post selected in the control register, the state after the second CNOT is $\ket{c^{(0)}_{b},w_{t}}_{2}=\ket{00}+\ket{11}+\ket{02}+\ket{12}$; the state after the subsequent action of a controlled swap $cS$ on work register modes $0$ and $2$ is found to be equivalent to that after an uncontrolled swap $uS$ on the same modes: $cS_{0,2}\ket{c^{(0)}_{b},w_{t}}_{2} = I \otimes uS_{0,2}\ket{c^{(0)}_{b},w_{t}}_{2} = \ket{00}+\ket{10}+\ket{11}+\ket{02}$.  When the first qubit is projected onto $\ket{0}$, the second qubit component of the Fourier transform is applied by the Identity operation $I$ followed by the Hadamard so that the final state is $\ket{c^{(0)}_{b},w_{t}}_{fin}=\ket{00}+\ket{00}+\ket{01}+\ket{02}+\ket{10}-\ket{10}-\ket{11}+\ket{12}$.

The crucial role of quantum interference in the Fourier transform can now be seen: the probability of observing the $0$ term in the control register is boosted by constructive interference and is three times that of the probability of observing the $1$ term, which suffers destructive interference; these terms are the digits $00$ and $10$ in the final probability distribution.  Therefore, contrast between these terms degrades with decoherence in the Fourier transform.

When the first control qubit has been projected onto $\ket{1}$, the state after the second CNOT is $\ket{c^{(1)}_{b},w_{t}}_{2}=\ket{00}+\ket{11}-\ket{02}-\ket{12}$.  The subsequent action of a controlled swap is equivalent to that of an uncontrolled swap up to a phase flip on terms with a $1$ in the control register: $I \otimes uS_{0,2}\ket{c^{(1)}_{b},w_{t}}_{2} = \ket{00}-\ket{11}-\ket{02}+\ket{10}$.  The phase flip is undone at the second qubit component the Fourier transform by applying $-i$ rather than $i$ so that the final state is $\ket{c^{(1)}_{b},w_{t}}_{fin}=\ket{00}-i\ket{00}+i\ket{01}-\ket{02}+\ket{10}+i\ket{10}-i\ket{11}-\ket{12}$.  Here, the Fourier transform imparts phases such that there is an equal probability of observing $0$ or $1$ in the control register.  These terms are the digits $01$ and $11$ in the final probability distribution.

\end{widetext}

\begin{thebibliography}{27}
\expandafter\ifx\csname natexlab\endcsname\relax\def\natexlab#1{#1}\fi
\expandafter\ifx\csname bibnamefont\endcsname\relax
  \def\bibnamefont#1{#1}\fi
\expandafter\ifx\csname bibfnamefont\endcsname\relax
  \def\bibfnamefont#1{#1}\fi
\expandafter\ifx\csname citenamefont\endcsname\relax
  \def\citenamefont#1{#1}\fi
\expandafter\ifx\csname url\endcsname\relax
  \def\url#1{\texttt{#1}}\fi
\expandafter\ifx\csname urlprefix\endcsname\relax\def\urlprefix{URL }\fi
\providecommand{\bibinfo}[2]{#2}
\providecommand{\eprint}[2][]{\url{#2}}

\bibitem[{\citenamefont{Feynman}(1982)}]{fe-ijtp-82-467}
\bibinfo{author}{\bibnamefont{Feynman},~\bibfnamefont{R.~P.}}
\bibinfo{title}{Simulating Physics with computers.}
\textit{\bibinfo{journal}{Int. J. Theor. Phy.}} \textbf{\bibinfo{volume}{21}},
\bibinfo{pages}{467-488} (\bibinfo{year}{1982}).

\bibitem[{\citenamefont{Deutsch}(1985)}]{de-prsla-400-97}
\bibinfo{author}{\bibnamefont{Deutsch},~\bibfnamefont{D.}}
\bibinfo{title}{Quantum theory, the Church-Turing principle and the universal quantum computer.}
\textit{\bibinfo{journal}{Proc. R. Soc. Lond. A}} \textbf{\bibinfo{volume}{400}},
\bibinfo{pages}{97-117} (\bibinfo{year}{1985}).

\bibitem[{\citenamefont{Nielsen and Chuang}(2000)}]{nielsen}
\bibinfo{author}{\bibnamefont{Nielsen},~\bibfnamefont{M.~A.}} \bibnamefont{\&}
  \bibinfo{author}{\bibnamefont{Chuang},~\bibfnamefont{I.~L.}}
  \textit{\bibinfo{title}{Quantum Computation and Quantum Information}}
  (\bibinfo{publisher}{Cambridge University Press}, \bibinfo{year}{2000}).

\bibitem[{\citenamefont{Shor}(1994)}]{sh-conf-94-124}
\bibinfo{author}{\bibnamefont{Shor},~\bibfnamefont{P.~W.}}
\bibinfo{title}{in Algorithms for Quantum Computation: Discrete Logarithms and Factoring}
\bibinfo{note}{(ed. Goldwasser, S.)}
\bibinfo{pages}{124-134}
 \bibinfo{journal}{(\textit{Proc. 35th Annu. Symp. Foundations of Computer Science}, 1994)}.

\bibitem[{\citenamefont{Ladd et~al.}(2010)\citenamefont{Ladd, Jelezko,
  Laflamme, Nakamura, Monroe, and OBrien}}]{la-nat-464-45}
\bibinfo{author}{\bibnamefont{Ladd},~\bibfnamefont{T.~D.}} \bibnamefont{\etal}
\bibinfo{title}{Quantum computers.}
  \textit{\bibinfo{journal}{Nature}} \textbf{\bibinfo{volume}{464}}, \bibinfo{pages}{45-53}
  (\bibinfo{year}{2010}).

\bibitem[{\citenamefont{Vandersypen et~al.}(2001)\citenamefont{Vandersypen,
  Steffen, Breyta, Yannoni, Sherwood, and Chuang}}]{va-nat-414-883}
\bibinfo{author}{\bibnamefont{Vandersypen},~\bibfnamefont{L.~M.~K.}} \bibnamefont{\etal}
\bibinfo{title}{Experimental realization of Shor's quantum factoring algorithm using nuclear magnetic resonance.}
\textit{\bibinfo{journal}{Nature}}
  \textbf{\bibinfo{volume}{414}}, \bibinfo{pages}{883-887} (\bibinfo{year}{2001}).

\bibitem[{\citenamefont{Lu et~al.}(2007)\citenamefont{Lu, Browne, Yang, and
  Pan}}]{lu-prl-99-250504}
\bibinfo{author}{\bibnamefont{Lu},~\bibfnamefont{C.-Y.}},
  \bibinfo{author}{\bibnamefont{Browne},~\bibfnamefont{D.~E.}},
  \bibinfo{author}{\bibnamefont{Yang}},~\bibfnamefont{T.} \bibnamefont{\&}
  \bibinfo{author}{\bibnamefont{Pan},~\bibfnamefont{J.-W.}}
\bibinfo{title}{Demonstration of a Compiled Version of Shor's Quantum Factoring Algorithm Using Photonic Qubits.}
  \textit{\bibinfo{journal}{Phys. Rev. Lett.}} \textbf{\bibinfo{volume}{99}},
  \bibinfo{eid}{250504} (\bibinfo{year}{2007}).

\bibitem[{\citenamefont{Lanyon et~al.}(2007)\citenamefont{Lanyon, Weinhold,
  Langford, Barbieri, James, Gilchrist, and White}}]{la-prl-99-250505}
\bibinfo{author}{\bibnamefont{Lanyon},~\bibfnamefont{B.~P.}} \bibnamefont{\etal}
\bibinfo{title}{Experimental Demonstration of a Compiled Version of Shor's Algorithm with Quantum Entanglement.}
  \textit{\bibinfo{journal}{Phys. Rev. Lett.}} \textbf{\bibinfo{volume}{99}},
  \bibinfo{eid}{250505} (\bibinfo{year}{2007}).

\bibitem[{\citenamefont{Politi et~al.}(2009)\citenamefont{Politi, Matthews, and
  O'Brien}}]{po-sci-325-1221}
\bibinfo{author}{\bibnamefont{Politi},~\bibfnamefont{A.}},
  \bibinfo{author}{\bibnamefont{Matthews},~\bibfnamefont{J.~C.~F.}},
  \bibnamefont{and} \bibinfo{author}{\bibnamefont{O'Brien},~\bibfnamefont{J.~L.}}
  \bibinfo{title}{Shor's Quantum Factoring Algorithm on a Photonic Chip.}
  \textit{\bibinfo{journal}{Science}}
  \textbf{\bibinfo{volume}{325}}, \bibinfo{pages}{1221} (\bibinfo{year}{2009}).

\bibitem[{\citenamefont{Parker and Plenio}(2000)}]{pa-prl-85-3049}
\bibinfo{author}{\bibnamefont{Parker},~\bibfnamefont{S.}} \bibnamefont{\&}
 \bibinfo{author}{\bibnamefont{Plenio},~\bibfnamefont{M.~B.}}
\bibinfo{title}{Efficient Factorization with a Single Pure Qubit and logN Mixed Qubits.}
 \textit{\bibinfo{journal}{Phys. Rev. Lett.}} \textbf{\bibinfo{volume}{85}},
  \bibinfo{pages}{3049-3052} (\bibinfo{year}{2000}).

\bibitem[{\citenamefont{Mosca and Ekert}(1999)}]{mo-LNCS-1509-174}
\bibinfo{author}{\bibnamefont{Mosca},~\bibfnamefont{M.}} \bibnamefont{\&}
  \bibinfo{author}{\bibnamefont{Ekert},~\bibfnamefont{A.}}
  \bibinfo{title}{The hidden subgroup problem and eigenvalue estimation on a quantum computer}
  \bibinfo{pages}{174--188}
  \bibinfo{journal}{(\textit{Lecture Notes in Computer Science}, Vol. \textbf{1509}, Springer, 1999)}.

\bibitem[{Par()}]{ParkerPlenioNote}
\bibinfo{note}{While the motivation of Ref. \onlinecite{pa-prl-85-3049} was to
  show that efficient factoring remains possible in systems where the
  preparation of pure states is challenging, it has not been widely appreciated
  that a key step---recycling of the control qubit also described in Ref.
  \onlinecite{mo-LNCS-1509-174}---dramatically reduces the resource requirement
  in all implementations: the reduction in resources comes with only a
  polynomial cost in time, while maintaing exponential speed-up of the
  algorithm.}

\bibitem[{\citenamefont{Aspuru-Guzik et~al.}(2005)\citenamefont{Aspuru-Guzik,
  Dutoi, Love, and Head-Gordon}}]{as-sci-309-1704}
\bibinfo{author}{\bibnamefont{Aspuru-Guzik},~\bibfnamefont{A.}},
  \bibinfo{author}{\bibnamefont{Dutoi},~\bibfnamefont{A.~D.}},
  \bibinfo{author}{\bibnamefont{Love},~\bibfnamefont{P.~J.}} \bibnamefont{\&}
  \bibinfo{author}{\bibnamefont{Head-Gordon},~\bibfnamefont{M.}}
\bibinfo{title}{Simulated Quantum Computation of Molecular Energies.}
  \textit{\bibinfo{journal}{Science}} \textbf{\bibinfo{volume}{309}},
  \bibinfo{pages}{1704-1707} (\bibinfo{year}{2005}).

\bibitem[{\citenamefont{Lanyon et~al.}(2010)\citenamefont{Lanyon, Whitfield,
  Gillett, Goggin, Almeida, Kassal, Biamonte, Mohseni, Powell, Barbieri
  et~al.}}]{la-nchem-2-106}
\bibinfo{author}{\bibnamefont{Lanyon},~\bibfnamefont{B.~P.}} \bibnamefont{\etal} 
\bibinfo{title}{Towards quantum chemistry on a quantum computer.}
\textit{\bibinfo{journal}{Nature Chem.}}
  \textbf{\bibinfo{volume}{2}}, \bibinfo{pages}{106-111} (\bibinfo{year}{2010}).

\bibitem[{\citenamefont{Veis and Pittner}(2010)}]{ve-jcp-133-194106}
\bibinfo{author}{\bibnamefont{Veis},~\bibfnamefont{L.}} \bibnamefont{\&}
  \bibinfo{author}{\bibnamefont{Pittner},~\bibfnamefont{J.}}
\bibinfo{title}{Quantum computing applied to calculations of molecular energies: CH2 benchmark.}
  \textit{\bibinfo{journal}{The Journal of Chemical Physics}}
  \textbf{\bibinfo{volume}{133}}, \bibinfo{eid}{194106}
  (\bibinfo{year}{2010}).

\bibitem[{\citenamefont{Whitfield et~al.}(2011)\citenamefont{Whitfield,
  Biamonte, and Aspuru-Guzik}}]{wh-mp-109-735}
\bibinfo{author}{\bibnamefont{Whitfield},~\bibfnamefont{J.~D.}},
  \bibinfo{author}{\bibnamefont{Biamonte},~\bibfnamefont{J.}} \bibnamefont{\&}
  \bibinfo{author}{\bibnamefont{Aspuru-Guzik},~\bibfnamefont{A.}}
\bibinfo{title}{Simulation of Electronic Structure Hamiltonians Using Quantum Computers.}
  \textit{\bibinfo{journal}{Mol. Phys.}} \textbf{\bibinfo{volume}{109}},
  \bibinfo{pages}{735-750} (\bibinfo{year}{2011}).

\bibitem[{\citenamefont{Li et~al.}(2011)\citenamefont{Li, Yung, Chen, Lu,
  Whitfield, Peng, Aspuru-Guzik, and Du}}]{li-sr-1-735}
\bibinfo{author}{\bibnamefont{Li},~\bibfnamefont{Z.}} \bibnamefont{\etal}
\bibinfo{title}{Solving Quantum Ground-State Problems with Nuclear Magnetic Resonance.}
  \textit{\bibinfo{journal}{Sci. Rep.}} \textbf{\bibinfo{volume}{1}}, \bibinfo{pages}{88}
  (\bibinfo{year}{2011}).

\bibitem[{\citenamefont{Griffiths and Niu}(1996)}]{gr-prl-76-3228}
\bibinfo{author}{\bibnamefont{Griffiths},~\bibfnamefont{R.~B.}} \bibnamefont{\&}
\bibinfo{author}{\bibnamefont{Niu},~\bibfnamefont{C.-S.}}
\bibinfo{title}{Semiclassical Fourier Transform for Quantum Computation.}
  \textit{\bibinfo{journal}{Phys. Rev. Lett.}} \textbf{\bibinfo{volume}{76}},
  \bibinfo{pages}{3228-3231} (\bibinfo{year}{1996}).

\bibitem[{Fir()}]{FirstNote}
\bibinfo{note}{In fact the same is generally true of a control register based on $d$ dimensional systems (qudits) and orders that are a power of $d$, $r=d^{p}$.}

\bibitem[{Sec()}]{SecondNote}
\bibinfo{note}{Shor's algorithm is designed to work for even orders only, as the classical subroutine must calculate $\gcd(x^{\frac{r}{2}}\pm1, N)$.  However, for certain choices of square coprime $x$ and odd order, the algorithm works.  For example, in the case of $x=2^{2j}$ with $j$ an integer, a numerical simulation of Shor's algorithm for the first (odd) $4851$ composite $N$ (product of two primes) finds approximately $58\%$ of this class of coprimes produces an odd order; of these approximately $36\%$ successfully lead to non trivial factors of $N$ in the classical $\gcd$ subroutine.  Order finding for $N=21$ with $x=4$, resulting in $r=3$ is one such successful example.}

\bibitem[{\citenamefont{Beckman et~al.}(1996)\citenamefont{Beckman, Chari,
  Devabhaktuni, and Preskill}}]{compiled}
\bibinfo{author}{\bibnamefont{Beckman},~\bibfnamefont{D.}},
\bibinfo{author}{\bibnamefont{Chari},~\bibfnamefont{A.~N.}},
\bibinfo{author}{\bibnamefont{Devabhaktuni},~\bibfnamefont{S.}} \bibnamefont{\&}
\bibinfo{author}{\bibnamefont{Preskill},~\bibfnamefont{J.}}
\bibinfo{title}{Efficient networks for quantum factoring.}
\textit{\bibinfo{journal}{Phys. Rev. A}} \textbf{\bibinfo{volume}{54}},
  \bibinfo{pages}{1034-1063} (\bibinfo{year}{1996}).
  
\bibitem[{Thi()}]{ThirdNote}
\bibinfo{note}{Note that, while encoding in higher dimensions will reduce the number of required photons, encoding the entire work register in a single $2^{m}$ level system typically requires resources exponential in $m$.}

\bibitem[{\citenamefont{Prevedel et~al.}(2007)\citenamefont{Prevedel, Walther,
  Tiefenbacher, Bohi, Kaltenbaek, Jennewein, and Zeilinger}}]{pr-nat-445-65}
\bibinfo{author}{\bibnamefont{Prevedel},~\bibfnamefont{R.}} \bibnamefont{\etal}
\bibinfo{title}{High-speed linear optics quantum computing using active feed-forward.}
\textit{\bibinfo{journal}{Nature}} \textbf{\bibinfo{volume}{445}}, \bibinfo{pages}{65-69}
  (\bibinfo{year}{2007}).

\bibitem[{\citenamefont{Zhou et~al.}(2011)\citenamefont{Zhou, Ralph, Kalasuwan,
  Zhang, Peruzzo, Lanyon, and O'Brien}}]{Zhou+Ralph2010}
\bibinfo{author}{\bibnamefont{Zhou},~\bibfnamefont{X.-Q.}} \bibnamefont{\etal}
\bibinfo{title}{Adding control to arbitrary unknown quantum operations.}
\textit{\bibinfo{journal}{Nature. Commun.}}
  \textbf{\bibinfo{volume}{2}}, \bibinfo{pages}{413} (\bibinfo{year}{2011}).

\bibitem[{\citenamefont{Ralph et~al.}(2001)\citenamefont{Ralph, Langford, Bell,
  and White}}]{ra-pra-65-062324}
\bibinfo{author}{\bibnamefont{Ralph},~\bibfnamefont{T.~C.}},
\bibinfo{author}{\bibnamefont{Langford},~\bibfnamefont{N.~K.}},
\bibinfo{author}{\bibnamefont{Bell},~\bibfnamefont{T.~B.}} \bibnamefont{\&}
\bibinfo{author}{\bibnamefont{White},~\bibfnamefont{A.~G.}}
\bibinfo{title}{Linear optical controlled-NOT gate in the coincidence basis.}
\textit{\bibinfo{journal}{Phys. Rev. A}} \textbf{\bibinfo{volume}{65}},
\bibinfo{pages}{062324} (\bibinfo{year}{2001}).

\bibitem[{\citenamefont{Hofmann and Takeuchi}(2001)}]{ho-pra-66-024308}
\bibinfo{author}{\bibnamefont{Hofmann},~\bibfnamefont{H.~F.}} \bibnamefont{\&}
\bibinfo{author}{\bibnamefont{Takeuchi},~\bibfnamefont{S.}}
\bibinfo{title}{Quantum phase gate for photonic qubits using only beam splitters and postselection.}
\textit{\bibinfo{journal}{Phys. Rev. A}} \textbf{\bibinfo{volume}{66}},
  \bibinfo{pages}{024308} (\bibinfo{year}{2001}).

\bibitem[{\citenamefont{O'Brien et~al.}(2003)\citenamefont{O'Brien, Pryde,
  White, Ralph, and Branning}}]{ob-nat-426-264}
\bibinfo{author}{\bibnamefont{O'Brien},~\bibfnamefont{J.~L.}},
\bibinfo{author}{\bibnamefont{Pryde},~\bibfnamefont{G.~J.}},
\bibinfo{author}{\bibnamefont{White},~\bibfnamefont{A.~G.}},
\bibinfo{author}{\bibnamefont{Ralph},~\bibfnamefont{T.~C.}} \bibnamefont{\&}
\bibinfo{author}{\bibnamefont{Branning},~\bibfnamefont{D.}}
\bibinfo{title}{Demonstration of an all-optical quantum controlled-NOT gate.}
\textit{\bibinfo{journal}{Nature}} \textbf{\bibinfo{volume}{426}},
  \bibinfo{pages}{264} (\bibinfo{year}{2003}).

\bibitem[{\citenamefont{O'Brien et~al.}(2004)\citenamefont{O'Brien, Pryde,
  Gilchrist, James, Langford, Ralph, and White}}]{ob-prl-93-080502}
\bibinfo{author}{\bibnamefont{O'Brien},~\bibfnamefont{J.~L.}} \bibnamefont{\etal}
\bibinfo{title}{Quantum Process Tomography of a Controlled-NOT Gate.}
\textit{\bibinfo{journal}{Phys. Rev. Lett.}} \textbf{\bibinfo{volume}{93}},
  \bibinfo{eid}{080502} (\bibinfo{year}{2004}).

\bibitem[{\citenamefont{Kwiat et~al.}(1999)\citenamefont{Kwiat, Waks, White,
  Appelbaum, and Eberhard}}]{kw-pra-60-773}
\bibinfo{author}{\bibnamefont{Kwiat},~\bibfnamefont{P.~G.}},
\bibinfo{author}{\bibnamefont{Waks},~\bibfnamefont{E.}},
\bibinfo{author}{\bibnamefont{White},~\bibfnamefont{A.~G.}},
\bibinfo{author}{\bibnamefont{Appelbaum},~\bibfnamefont{I.}} \bibnamefont{\&}
\bibinfo{author}{\bibnamefont{Eberhard},~\bibfnamefont{P.~H.}}
\bibinfo{title}{Ultrabright Source of Polarization-Entangled Photons.}
\textit{\bibinfo{journal}{Phys. Rev. A}}
\textbf{\bibinfo{volume}{60}}, \bibinfo{pages}{R773-R776} (\bibinfo{year}{1999}).
  
\bibitem[{Fou()}]{FourthNote}
\bibinfo{note}{All fidelities are calculated as $1- trace \,distance$.  The \emph{trace distance} between two probability distributions $\{p_{x}\}$ and $\{q_{x}\}$ (sometimes known as the $L_{1}$ \emph{distance} or \emph{Kolmogorov distance}) is defined \cite{nielsen} by the equation $D(p_{x},q_{x})\equiv \frac{1}{2} \sum_{x} | p_{x} - q_{x} |$.  Error bars on bar charts represent the propagated experimental uncertainty arising from Poissonian counting statistics.}

\bibitem[{FouB()}]{FourthNoteB}
\bibinfo{note}{Error bars on these data points represent experimental uncertainty arising from Poissonian counting statistics.}

\bibitem[{Fif()}]{FifthNote}
\bibinfo{note}{Fidelity between two quantum states $\rho$ and $\sigma$ is defined\cite{nielsen} to be $F(\rho,\sigma)\equiv \tr \sqrt{\rho^{1/2} \sigma \rho^{1/2}}$.}

\bibitem[{\citenamefont{Knill et~al.}(2001)\citenamefont{Knill, Laflamme, and
  Milburn}}]{kn-nat-409-46}
\bibinfo{author}{\bibnamefont{Knill},~\bibfnamefont{E.}},
\bibinfo{author}{\bibnamefont{Laflamme}},~\bibfnamefont{R.} \bibnamefont{\&}
\bibinfo{author}{\bibnamefont{Milburn},~\bibfnamefont{G.~J.}}
\bibinfo{title}{A scheme for efficient quantum computation with linear optics.}
\textit{\bibinfo{journal}{Nature}} \textbf{\bibinfo{volume}{409}}, \bibinfo{pages}{46-52}
(\bibinfo{year}{2001}).

\end{thebibliography}
\end{document}